\documentclass[twocolumn, showpacs,aps,prl,letterpaper,groupeaddress,nofootinbib, tightenlines]{revtex4-1}

\usepackage{graphicx}
\usepackage{dcolumn}
\usepackage{bm}
\usepackage{mciteplus}
\usepackage{amsmath}
\usepackage{amsfonts,amssymb}
\usepackage{color}
\usepackage{cancel}
\usepackage{enumerate,soul}

\begin{document}

\title{On Emergent Gravity, Black Hole Entropy and Galactic Rotation Curves}

\author{I. D\'iaz-Salda\~na$^{1}$}
\author{J. C. L\'opez-Dom\'inguez$^{2}$}
\author{M. Sabido$^{1,3}$}
\affiliation{$^{1}$Departamento  de F\'{\i}sica de la Universidad de Guanajuato,\\
 A.P. E-143, C.P. 37150, Le\'on, Guanajuato, M\'exico.}
\affiliation{$^{2}$Unidad Acad\'emica de F\'isica, Universidad Aut\'onoma de Zacatecas, Calzada Solidaridad esquina con Paseo a la Bufa S/N C.P. 98060, Zacatecas, M\'exico.}
\affiliation{$^3$ Department of Theoretical Physics, University of the Basque Country UPV/EHU, P.O BOX 644, 48080 Bilbao, Spain.}

\begin{abstract}
In this work we derive a generalized Newtonian gravitational force and show that it can account for the anomalous galactic rotation curves. We derive the entropy-area relationship applying the Feynman-Hibbs procedure  to the supersymmetric Wheeler-DeWitt equation of the Schwarzschild black hole. We obtain the modifications to the Newtonian gravitational force from the entropic formulation of gravity.
\end{abstract}

\draft
\maketitle
\section{Introduction} \label{Int}
The logical structure of General Relativity (GR) is one of the greatest achievements in theoretical physics. The current results from gravitational wave astronomy cements GR as the appropriate theory to describe the gravitational interaction. Although the open problem of dark energy and dark matter is compatible with GR (if one proposes new exotic sources of matter and energy), current observations do not discard alternative theories of gravity. Furthermore, considering that after many decades of research a complete quantum theory of gravity is still missing, one can strongly look at alternatives to GR. We can follow the traditional line of reasoning by considering gravity as a fundamental interaction and from some fundamental principle write down the corresponding theory (i.e. $f(R)$ \cite{Sotiriou:2008rp}, massive gravity \cite{deRham:2014zqa}, Horndeski, etc.). Another approach is to consider gravity  not as fundamental interaction but as an emergent phenomenon \cite{Jacobson:1995ab}.
During this decade there has been a renewed interest in this idea. It started with Verlinde's ideas presented in \cite{Verlinde:2010hp}, where he claims that Newtonian gravity is an entropic force, in the sense of the emergent forces that are present in the study of polymers. This approach is motivated from the ideas in holography together with the area relation for the entropy of black holes. Since in this formulation Newtonian gravity has an entropic origin, one can propose modifications to Newtonian gravity by analyzing modifications to the entropy area law \cite{Modesto:2010rm,Martinez-Merino:2017xzn}.

Research on black hole physics has uncovered several mysteries: Why is the statistical
black hole entropy proportional to the horizon area?, What happens to the information in black
hole evaporation?. The answers to these questions and more others have been extensively studied in the
literature, however, a definitive answer to the microscopic description of black holes has yet to be found. 
For supersymmetric theories there is a bound on the mass $M$ of the states related to the  supercharges $Q$, it is known as the BPS bound $M\ge Q$. The mass of the states that saturate this bound (BPS states) are protected and do not receive quantum corrections. Charged black holes that satisfy $M=Q$ are called extremal black holes, which turn out to have no Hawking temperature, are quantum mechanically stable and their microscopic description is exact.
Unfortunately, this approach {can not be applied to} the Schwarzschild black hole. In \cite{LopezDominguez:2009ue,LopezDominguez:2011zz}, the authors propose a supersymmetric generalization to the Schwarzschild and Schwarzschild-(anti)de Sitter black holes. Their approach begins with the known diffeomorphism between the Kantowski Sachs (KS) cosmological solution and the Schwarzschild black hole solution. For this model they find the Wheeler-DeWitt equation (WDW) and show that  after a WKB approximation the black hole solution is recovered. Then, using the methods of supersymmetric quantum cosmology \cite{Graham:1991av,Obregon:1999wt},
a supersymmetric generalization of the Schwarzschild black hole can be constructed. In \cite{Obregon:2000zd},
the authors use the Feynman-Hibbs path integral procedure \cite{feynman} to calculate the temperature and entropy of the Schwarzschild black hole. This approach allows us to incorporate quantum
corrections to the partition function through the {\it
``corrected"} potential and has been successfully applied to the study of the thermodynamics of different black hole models \cite{LopezDominguez:2006wd,Bastos:2009ae}. 
Using this method, we calculate the entropy-area relationship for the supersymmetric generalization of the Schwarzschild black hole. Furthermore, using {Verlinde's proposal} we find the corrections to Newton's law of gravitation and show that the contributions to Newtonian gravity give the correct rotation curves for galaxies. 

\section{The supersymmetric WDW equation for Schwarzschild metric} \label{difeo}
In order to obtain the supersymmetric generalization of the WDW equation for the Schwarzschild black hole we will use its relationship with the KS model. The
Schwarzschild black hole is described by the metric
\begin{eqnarray}
ds^{2}&=&-\left ( 1-\frac{2m}{r} \right )dt^{2}+\left ( 1-\frac{2m}{r} \right )^{-1}dr^{2}\\
&+&r^{2}\left(d\theta^{2}+\sin^{2}\theta d\varphi^{2}  \right),\nonumber
\end{eqnarray}
for the case $r<2m$ the $g_{rr}$ and the $g_{tt}$ components of the metric change in sign and $\partial_{t}$ becomes a space-like vector, {hence} if we perform the transformation $t\leftrightarrow r$,
and compare with the Misner parametrization of the KS metric
we identify    
\begin{equation}
N^{2}=\left(\frac{2m}{t}-1\right)^{-1}, e^{2\sqrt{3}\xi }=\frac{2m}{t}-1, e^{-2\sqrt{3}\left ( \xi +\Omega \right )}=t^{2},\label{iden}
\end{equation}
this establishes the diffeomorphism with KS. Using this result in the Einstein-Hilbert action
and performing an integration over the spatial coordinates, we get an effective Lagrangian from which it is straight forward to obtain the Hamiltonian constraint
\begin{equation}
H=p_{\xi}^{2}-p_{\Omega}^{2}-48e^{-2\sqrt{3}\Omega}.\label{hamilton}
\end{equation}
From standard canonical quantization, with the usual identifications for the canonical momenta $p_{\xi}=-i\frac{\partial }{\partial \xi}$, $p_{\Omega}=-i\frac{\partial }{\partial \Omega}$, the WDW equation derived from the Hamiltonian constraint is
\begin{equation}
\left ( -\frac{\partial^2 }{\partial \Omega ^2}+\frac{\partial^2 }{\partial \xi ^2}+48e^{-2\sqrt{3}\Omega } \right )\Psi (\Omega ,\xi )=0,
\label{wdwks}
\end{equation}
this equation has been used to study quantum black holes \cite{ryan}. Using the plane wave solution for the variable $\xi$, the WDW equation takes the form
\begin{equation}
\left ( -\frac{d^{2}}{d\Omega^{2}} +48e^{-2\sqrt{3}\Omega }\right )\chi(\Omega)=3\nu^{2}\chi(\Omega).\label{QE}
\end{equation}
We now proceed to obtain the supersymmetric version of the WDW equation Eq.\eqref{wdwks}, for this purpose we will follow \cite{Graham:1991av}. 
For homogeneous models the Hamiltonian $H_{0}$ can be written as 
\begin{equation}
2H_{0}=\mathcal{G}^{\mu \nu }p_{\mu }p_{\nu }+\mathcal{U}(q^{\mu}),\label{H022}
\end{equation}
 where $q^{\mu }$ are the minisuperspace coordinates, $\mathcal{G}^{\mu \nu }$ is the minisuperspace metric and $\mathcal{U}(q^{\mu})$ is the potential. Also, it is possible to find a function $\Phi(q^{\nu})$ satisfying 
\begin{equation}\mathcal{G}^{\mu \nu }\frac{\partial \Phi }{\partial q^{\mu }}\frac{\partial \Phi }{\partial q^{\nu }}=\mathcal{U}(q^{\alpha}).\label{ec}\end{equation}
To construct the supersymmetric Hamiltonian first we need the 
supercharges 
\begin{equation}
\mathcal{Q}=\psi^{\mu } \left ( p_{\mu } +i\frac{\partial \Phi }{\partial q^{\mu }}\right ),\quad \bar{\mathcal{Q}}=\bar{\psi }^{\mu }\left ( p_{\mu }-i\frac{\partial \Phi}{\partial q^{\mu }} \right ), \label{spcosm}
\end{equation}
where $\bar{\psi }^{\mu}$, and $\psi^{\nu }$ are  Grassmann variables and satisfy the algebra
\begin{equation}\left \{ \bar{\psi }^{\mu } ,\bar{\psi }^{\nu } \right \}=\left \{\psi ^{\mu},\psi ^{\nu }  \right \}=0, \quad \left \{ \bar{\psi }^{\mu },\psi ^{\nu } \right \}=\mathcal{G}^{\mu \nu }.\label{alg}\end{equation}
The supersymmetric Hamiltonian  is obtained from the algebra of the supercharges $2H_{S}=\{\mathcal{Q},\bar{\mathcal{Q}}\}$, this gives
\begin{equation}
2H_{S}=\mathcal{G}^{\mu \nu }p_{\mu }p_{\nu }+\mathcal{U}(q^{\mu})+\frac{\partial ^{2}\Phi }{\partial q^{\mu }\partial q^{\nu }}\left [ \bar{\psi }^{\mu},\psi^{\nu }  \right ].\label{hsusycosm}
\end{equation}
This supersymmetric Hamiltonian is the supersymmetric generalization of Eq.(\ref{H022}), it is the sum of the ``bosonic'' Hamiltonian and the contribution $\frac{\partial ^{2}\Phi }{\partial q^{\mu }\partial q^{\nu }}$. 
This Hamiltonian is fully determined once we adopt a suitable representation of the Grassmann variables
\begin{equation}
\begin{aligned}
&\psi ^{\xi}=\left(\begin{smallmatrix}
0 &0  &0  &0 \\ 
1 &0  &0  &0 \\ 
 0&0  &0  &0 \\ 
 0&0  &\text{-}1  &0 
\end{smallmatrix}\right),
 \quad \bar{\psi }^{\xi}=\left(\begin{smallmatrix}
0 &1  &0  &0 \\ 
 0&0  &0  &0 \\ 
 0&0  &0  &\text{-}1 \\ 
 0&0  &0  &0 
\end{smallmatrix}\right),\\
&\psi ^{\Omega}=\left(\begin{smallmatrix}
0 &0  &0  &0 \\ 
0&0  &0  &0 \\ 
 1&0  &0  &0 \\ 
 0&1  &0  &0 
\end{smallmatrix}\right), \quad \bar{\psi }^{\Omega}=\left(\begin{smallmatrix}
0 &0  &\text{-}1 &0 \\ 
 0&0  &0  &\text{-}1 \\ 
 0&0  &0  &0 \\ 
 0&0  &0  &0 
\end{smallmatrix}\right).
\end{aligned}
\end{equation}
{From Eq.\eqref{wdwks} we identify $\mathcal{U} (\Omega,\xi)=48e^{-2\sqrt{3}\Omega }$ and from Eq.\eqref{ec} we obtain a differential equation for $\Phi$ whose solution is given by
\begin{align}
\Phi=&-4\left [ \sqrt{e^{-2\sqrt{3}\Omega }+\epsilon ^{-2/3}}-\epsilon ^{-1/3}{\rm arcsinh}\left (\epsilon ^{-1/3} e^{\sqrt{3}\Omega } \right ) \right ]\nonumber\\
&+4\sqrt{3}\epsilon^{-1/3}\xi,
 \end{align}
with $\epsilon={\rm constant}$. Since $\left[ \bar{\psi }^{\Omega},\psi^{\Omega } \right]={\rm diag}(-1,-1,1,1)$, the supersymmetric Hamiltonian will have two independent components which only differ in the sign of the modified potential. Now the proposed WDW equation for the supersymmetric quantum Schwarzschild black hole is 
\begin{align}\label{susyec}
&\left[ -\frac{\partial^2}{\partial\Omega^2}+\frac{\partial^2}{\partial\xi^2}\right. \\
&\left. +12\left( 4\pm\frac{1}{\sqrt{e^{-2\sqrt{3}\Omega}+\epsilon^{-2/3}}} \right)e^{-2\sqrt{3}\Omega} \right] \Psi_{\pm}^{S}(\Omega, \xi)=0.\nonumber
\end{align}
The wave function has four components, although only two are linearly independent, also the contributions of supersymmetry are encoded in the modified potential. Finally, it is worth mentioning  that  Eq.(\ref{wdwks}) is recovered from Eq.(\ref{susyec}), by taking the limit $\epsilon\to 0$.

\section{Modified  Entropy-Area relationship} \label{BHT}
Let us start by reviewing the calculation of the entropy for the Schwarzschild black hole, using the Feynman-Hibbs procedure. This approach was originally applied to the Schwarzschild  black hole \cite{Obregon:2000zd} and subsequently used for different black hole models \cite{LopezDominguez:2006wd,Bastos:2009ae,mena}. 
In the limit {of} small $\Omega$ and taking $x=l_{P}(\sqrt{6}\Omega-1/\sqrt{2})$, Eq.(\ref{QE}) can be written as
\begin{equation}
\left ( -\frac{1}{2}l_{P}^{2}E_{P}\frac{d^{2}}{dx^{2}}+4\frac{E_{P}}{l_{P}^{2}}x^{2} \right )\chi (x)= E_{P}\left ( \frac{\nu ^{2}}{4}-2 \right )\chi (x),\label{eccuan}
\end{equation}
we can see that Eq.(\ref{eccuan}) is the usual quantum harmonic oscillator if we identify $\hbar \omega=\sqrt{\frac{3}{2\pi}}E_p$ and $\frac{\hbar ^2}{m}=l_{P}^{2}E_{P}$. 

\noindent To compute the ``corrected'' partition function of the black hole we apply the Feynman-Hibbs procedure. This approach is based on exploiting the similarities of the expression of the density matrix and the kernel of Feynman's path integral approach to quantum mechanics. By doing a Wick rotation $t\to i\beta$, we get the Boltzmann factor and the  kernel is transformed to the density matrix. The kernel is calculated along  the paths that go from $x_1$ to $x_2$, if we consider small $\Delta t$ (small $\beta$). Then, when calculating the partition function, only the paths that stay near $x_1$ have a non-negligible contribution (the exponential in the expression for the density matrix gives a negligible contribution to the sum from the other paths). Therefore, the potential to a first-order approximation can be written as $V(x)\approx V(x_1)$ for all the contributing paths.
 In this approximation, we can formally establish a map from the path integral formulation of quantum mechanics to the classical canonical partition function. To introduce quantum effects, we must incorporate the changes to the potential along the path; in particular, we are interested in the first-order effects. For this, we start by doing a Taylor expansion around the mean position $\tilde x$ along any path. Calculating the kernel with $\tilde x$ and doing the Wick rotation, we get the modified partition function. This partition function is calculated in a classical manner, but with the corrected potential, which is a mean value of the potential $V(x)$ averaged over points near $\tilde x$ with a Gaussian distribution. Therefore, to calculate the partition function with quantum corrections \cite{Obregon:2000zd,LopezDominguez:2006wd,Bastos:2009ae}, we use the corrected potential. 
According to this procedure, the  corrected partition function is 
\begin{equation}
Z=\sqrt{\frac{m}{2\pi\beta\hbar^2}}\int^{\infty}_{-\infty}e^{-\beta U(\tilde{x})}d\tilde{x},\label{fundep}
\end{equation}
where $\beta=1/k_{B}T$ and $U(\tilde{x})$ is the corrected potential given by
\begin{equation}
U(\tilde{x})=\sqrt{\frac{12m}{2\pi \beta \hbar^{2}}}\int_{-\infty}^{\infty}V(\tilde{x}+y)e^{-6 y^{2}m/\beta \hbar^{2}}dy.\label{potc}
\end{equation} 
Now we substitute the potential of the WDW equation in Eq.\eqref{fundep} and Eq.\eqref{potc}, this gives the corrected partition function
\begin{equation}
Z=\sqrt{\frac{2\pi}{3}}\frac{1 }{\beta E_{P}}e^{-\beta ^{2}E_{P}^{2}/16\pi}.\label{FP}
\end{equation}
From the partition function it is straightforward to calculate the temperature and entropy for the Schwarzschild black hole.
We begin with the internal energy  
\begin{equation}
{E}=-\frac{\partial}{\partial \beta}\ln Z,\label{EI2}
\end{equation}
{which gives a relation between the black hole corrected temperature $\beta$ and its mass $M$}. In terms of Hawking temperature $\beta_{H}=\frac{8\pi Mc^{2}}{E_{P}^{2}}$, the corrected temperature of the black hole is
\begin{equation}
\beta=\beta_{H}\left ( 1-\frac{1}{\beta _{H}}\frac{1}{Mc^{2}} \right ),\label{temp}
\end{equation} 
where we can observe an extra contribution to the temperature proportional to $\beta_{H}^{-1}$.

\noindent To calculate the entropy we use
\begin{equation}\label{relacion}
\frac{S}{k_{B}}=\ln{Z}+\beta {E},
\end{equation}
by relating the Bekenstein-Hawking entropy to the Hawking temperature as $Mc^{2}\beta_{H}=2\frac{S_{BH}}{k_{B}}$ we get entropy
\begin{equation}
\frac{S}{k_{B}}=\frac{S_{BH}}{k_{B}}-\frac{1}{2}\ln \frac{S_{BH}}{{k_{B}}}+\mathcal{O}(S_{BH}^{-1}).\label{entrop}
\end{equation}
This result has the interesting feature that the logarithmic correction agrees with the one obtained in string theory as well as in loop quantum gravity  \cite{Domagala:2004jt,Mukherji:2002de,Sen:2012dw}. 

Now we apply this method to obtain the temperature and entropy of the supersymmetric Schwarzschild black hole. Since the potential in Eq.\eqref{susyec} depends only on the $\Omega$ coordinate, we use the plane wave solution for $\xi$. Following the same procedure and similar approximations as in the original case, the equation  takes the form
\begin{equation}\left [ -\frac{1}{2}l_{P}^{2}E_{P}\frac{d^{2}}{dx^{2}}+4\frac{E_{P}}{l_{P}^{2}}x^{2}\pm\frac{1}{2}\frac{E_{P}}{l_{P}^{4}}\epsilon x^{4} \right ]\chi_{\pm}^{S} (x)=\eta^{2}\chi_{\pm}^{S} (x). \label{ecaprox}
\end{equation}
From now on we will take the  $(-)$ case, the $(+)$ case follows straightforwardly by transforming $\epsilon\to-\epsilon$.
Now we apply the Feynman-Hibbs procedure for the potential of the form $V(x)=\frac{m\omega^2}{2}x^2+\lambda x^4$. Following the bosonic case a straightforward calculation gives the  corrected partition function for the supersymmetric model
\begin{eqnarray}\label{FPSUSY}
Z_{S}&=&\sqrt{\frac{2 \pi}{3}}\frac{1}{\beta E_{P}}\exp{\left ( -\frac{\beta^{2}E_{P}^{2}}{16\pi}-\frac{\beta^{3}E_{P}^{3}\epsilon}{96} \right )}\\
&&\times \left ( 1+\frac{\pi \beta E_{P} \epsilon}{3} \right )^{-1/2}.\nonumber
\end{eqnarray}
As before, this partition function reduces to the original one in the limit $\epsilon\to 0$. 
For temperature we proceed as before, using the partition function Eq.\eqref{FPSUSY} the internal energy is
\begin{equation}
\frac{1}{\beta}+\frac{E_{P}^{2}\beta}{8 \pi}+\frac{E_{P}^{3}\beta^{2}}{32}\epsilon+\frac{\pi E_{P}}{6}\epsilon-\frac{\pi^{2}E_{P}^{2}\beta}{18}\epsilon^{2}=Mc^2, \label{ec1}
\end{equation}
solving for $\beta$ in terms of the Hawking temperature gives 
\begin{equation}\label{Tsusy}
\beta=\beta_{H}\left [ 1-\frac{1}{\beta _{H}Mc^{2}}+f(\epsilon ) \frac{1}{\left ( \beta _{H}Mc^{2} \right )^{1/2}}\right ],
\end{equation}
where $f(\epsilon)= \frac{2}{3}\epsilon^{3}-\epsilon$, and for convenience, the parameter $\epsilon$ has been redefined by  $\epsilon \rightarrow \frac{2^{1/2}\pi ^{3/2}}{3}\epsilon$. We see the supersymmetric contribution to the temperature is proportional to $\beta_{H}^{-1/2}$, also in the limit $\epsilon \to 0$ we recover {Eq. \eqref{temp}}.

\noindent From the partition function Eq.\eqref{FPSUSY} and the temperature, the entropy for the supersymmetric model is given by
\begin{equation}
\frac{S}{k_{B}}=-\frac{1}{2}\ln \frac{3E_{P}^{2}\beta ^{2}}{2\pi }+\frac{E_{P}^{2}\beta ^{2}}{16\pi }+\frac{E_{P}^{3}\beta ^{3}}{48}\epsilon -\frac{\pi ^{2}E_{P}^{2}\beta ^{2}}{18}\epsilon^{2}+1.
\end{equation}
Solving in terms of the Bekenstein-Hawking entropy $S_{BH}/k=A/4l_P^2$  we arrive  to the entropy-area relationship
\begin{eqnarray}
\frac{S[A]}{k_B}&=&\frac{\left [1+\Delta(\epsilon) \right ]}{4l_{P}^{2}}A-\frac{1}{2}\ln{\frac{A}{4l_{P}^{2}}}\nonumber\\
&+&\Gamma(\epsilon)\left(\frac{A}{4l_{P}^{2}}\right)^{1/2}+\epsilon\left(\frac{A}{2l_{P}^{2}}\right)^{3/2},\label{entr}
\end{eqnarray}
where $\Gamma(\epsilon)=2^{1/2}f(\epsilon )-3\cdot 2^{1/2}\epsilon +3\cdot 2^{1/2}\epsilon f^{2}(\epsilon )-2^{5/2}\epsilon ^{2}f(\epsilon )$ is and odd function of $\epsilon$ and $\Delta(\epsilon)=6\epsilon f(\epsilon )-4\epsilon ^{2}$ is an even function.\\
The modifications to the entropy of the proposed supersymmetric Schwarzschild model can be understood as follows.
The first term is the usual Bekenstein-Hawking entropy, the logarithmic correction seems to be a universal correction and has been  derived in different approaches in the study of black holes \cite{Obregon:2000zd,Domagala:2004jt,Mukherji:2002de,Sen:2012dw}. The third term is proportional to a linear length, this term can be interpreted as an effective contribution due to a self-gravitating gas,  this behavior in the scaling is consistent with the entropy calculated for a self-gravitating gas where  $S\sim V^{1/3}$  \cite{deVega:2001zk}. Finally the last term is proportional to the volume and is the usual behavior that the entropy has for non gravitational systems, and corresponds to effective volumetric short distance interactions. This type of volumetric modification, in the context of emergent gravity, was first studied in \cite{Modesto:2010rm}, where the authors justify the introduction of this term by using arguments from 
loop quantum gravity.

\section{Emergent Modified Newtonian Gravity}\label{eg}
The relation between entropy and gravity to derive Einstein's equations was put forward by Jacobson \cite{Jacobson:1995ab}. In more recent works Verlinde showed a relation between the Entropic force and the Newtonian gravity. He proposed that gravity is an effective force which emerges from the entropy, such as the emergent entropic forces in polymers. These forces are connected to the entropy via the thermodynamic equation of state $F\Delta x=T \Delta S$.
Verlinde's approach relates the entropy with the information contained in a surface $\mathcal{S}$ that surrounds a mass $M$, which is very near from a test mass $m$, from which the entropic derivation of Newtonian gravity is obtained \cite{Verlinde:2010hp}. These ideas have inspired several models that attempt to modify Newtonian gravity. This has been achieved by proposing quantum modification to the entropy  or by  using new definitions of entropy \cite{Martinez-Merino:2017xzn} to find modifications to Newton's gravitational force and in some cases modifications to gravity in the cosmological scenario \cite{Sheykhi:2010yq}. We can write a generic modification to the entropy-area relationship  as
\begin{equation}
\frac{S}{k_{B}}=\frac{A}{4l_{P}^{2}}+\mathfrak{s}(A),\label{moden}
\end{equation}
the first term corresponds to the usual area law and the second term, $\mathfrak{s}(A)$, includes the other contributions to the entropy.Using Verlinde's entropic Newtonian gravity, modifications to Newton's law of gravitation \cite{Modesto:2010rm} can be obtained from
\begin{equation}\label{verlinde} 
\mathbf{F}_{M}=-\frac{GMm}{R^{2}}\left[1+4l_{P}^{2}\frac{\partial \mathfrak{s}}{\partial A}\right]_{A=4\pi R^{2}}\mathbf{\hat{R}}.
\end{equation}
With this in mind we will find the emergent modified Newtonian force related with the area-entropy derived in the previous section. In order to proceed, we separate the entropy of the supersymmetric model in accordance to Eq.\eqref{moden}. The calculation for the entropic Newtonian force is straightforward
 \begin{eqnarray}
\mathbf{F}_{M}&=&-\frac{G_{eff}Mm}{R^{2}}\left[1+\frac{3\sqrt{2\pi}  }{l_{P}(1+\Delta(\epsilon))} \epsilon R \right.\\
&+&\left.\frac{l_{P}\Gamma(\epsilon)}{2\sqrt{\pi}(1+\Delta(\epsilon))}\frac{1}{R}
-\frac{l_{P}^{2}}{2\pi(1+\Delta(\epsilon))}\frac{1}{R^2}\right]\mathbf{\hat{R}},\nonumber \label{modforce}
\end{eqnarray}
where $G_{eff}=(1+\Delta(\epsilon))G$ is the {effective} gravitational constant.  
We can derive the modified gravitational potential by integrating  Eq.\eqref{modforce}, which up to a arbitrary integration constant $\sigma$ is
\begin{eqnarray}
\Phi_{M}&=&-\frac{G_{eff}M}{R}\left[1-\frac{3\sqrt{2\pi}}{l_{P}(1+\Delta(\epsilon))} \epsilon R\ln{\frac{R}{\sigma}}\right.\\
&+&\left.\frac{l_{P}\Gamma(\epsilon)}{4\sqrt{\pi}(1+\Delta(\epsilon))}\frac{1}{R}-\frac{l_{P}^{2}}{6\pi(1+\Delta(\epsilon))}\frac{1}{R^{2}}\right].\nonumber
\end{eqnarray}
It is also straightforward  to obtain an effective  matter density such that the Poisson equation $\nabla ^{2}\Phi_{M}=4\pi\rho _{eff}$ is satisfied
\begin{equation}
\rho_{eff}=\frac{G M}{2\pi R}\frac{l^2_P}{2\pi R^4}\left[1-\frac{\sqrt{\pi}}{2 l_P}\Gamma(\epsilon) R+\frac{3}{2}\left(\frac{\sqrt{2\pi}}{l_P}\right)^3\epsilon R^3\right].
\end{equation}
We can have different interpretations to this modified matter density, one can consider that the origin of $\rho_{eff}$ is a consequence  to the presence of a point mass $M$ and through some unknown process gives $\rho_{eff}$. The other possibility is to consider that we have a point particle of mass $M$  that generates the potential $\Phi_M$ by using the appropriate limit of some unknown theory of gravity.

\noindent If we consider a particle of mass $m$ in a circular orbit of radius $R$, according to Eq.\eqref{modforce}, the velocity of the particle is given by
\begin{eqnarray}\label{velocidad2}
\frac{mv^2}{R}&=&\frac{G_{eff}Mm}{R^{2}}\left[1+\frac{3\sqrt{2\pi}  }{l_{P}(1+\Delta(\epsilon))} \epsilon R \right.\\
&+&\left.\frac{l_{P}\Gamma(\epsilon)}{2\sqrt{\pi}(1+\Delta(\epsilon))}\frac{1}{R}
-\frac{l_{P}^{2}}{2\pi(1+\Delta(\epsilon))}\frac{1}{R^2}\right].\nonumber
\end{eqnarray}
Such is the case of a star in circular motion around the center of a galaxy. The second term in Eq.\eqref{velocidad2}
is the leading order term for very large $R$ and corresponds to the entropic volumetric correction. At this point, we observe that our model has one free parameter $\epsilon$ which can be fit so we can reproduce the observed galactic rotations curves. On the other hand, the Modified Newtonian Dynamics (MOND), is a one parameter phenomenological theory, that reproduces the observed galactic rotation curves \cite{mond}. Of particular interest is the behavior for large $R$, since in this limit the velocity obtained in our model is a constant, like the velocity one obtains from MOND in the same limit. We can exploit this fact in order to relate our parameter $\epsilon$ with the characteristic quantities of MOND. 
In the limit of large $R$ the velocity is given by
\begin{equation}
v^2\approx\frac{3\sqrt{2\pi}G M}{l_P}\epsilon,
\end{equation}
it is worth mentioning that the velocity depends on the regular gravitational constant $G$ and is linear on the parameter $\epsilon$}. Comparing with the velocity obtained for large $R$ in MOND, $v^2=\sqrt{G Ma_0}$, we can see that the results agree if
\begin{equation}
\epsilon=\frac{1}{6\pi}\sqrt{\frac{a_0h}{M c^3}}.
\end{equation}
{We can constrain the upper bound for $\epsilon$ by using the Planck mass $M_{P}$. Considering the characteristic acceleration of MOND $a_0\approx10^{-10}m/s^2$ then  $\epsilon < 10^{-30}$.}

\section{Conclusions and final remarks }\label{conclusiones}

The complexities of defining a quantum theory of gravity has led to consider gravity as an emergent phenomena. This opens a new paradigm for the understanding of the origin of the gravitational interaction. By assuming an entropic origin of  Newtonian gravity we can study new effects by considering modifications to the Bekenstein-Hawking entropy. 
In this work, 
we used the supersymmetric minisuperspace approach for the Schwarzschild black hole  to construct a supersymmetric generalization and from the Feynman-Hibbs approach we obtained the  entropy-area relationship. It is worth mentioning, that  except for the logarithmic term that has a quantum origin, the remaining terms are related to the supersymmetric modification. Assuming an entropic origin to gravity, we constructed a generalized Newtonian force,  the modified gravitational potential and the effective matter density. 
When considering the case of circular orbit for very large radius, this modified theory of gravity can account for the anomalous galaxy rotation curves. Therefore one can conjecture that if gravity is emergent, supersymmetry can substitute (under some conditions) the need for dark matter to explain the rotation curves. 
Although this results are encouraging,  further exploration is needed to establish this model as a replacement to dark matter.

\section*{Acknowledgements}
This work is supported by CONACyT grants 257919, 258982. M. S. is supported by the CONACyT program  ``Estancias sab\'aticas en el extranjero'', grant 31065. J.C.L-D. is supported by UAZ grant UAZ-2016-37235. I. D-S. thanks CONACyT support.
 
\end{document}